\documentclass[reprint,unsortedaddress,amsmath,amssymb,prl]{revtex4-2}

\usepackage{graphicx}
\usepackage{dcolumn}
\usepackage[mathlines]{lineno}
\usepackage{ulem}
\usepackage{xcolor}
\begin{document}

\author{Manobina Karmakar}
\affiliation{Department of Physics, Indian Institute of Technology Kharagpur,  Kharagpur, India 721302}
\author{Sayantan Bhattacharya}
\affiliation{Department of Physics, Indian Institute of Technology Kharagpur,  Kharagpur, India 721302}
\affiliation{\textit{presently at} Department of Chemistry, University of Sheffield, Sheffield S3 7HF, United Kingdom}
\author{Subhrajit Mukherjee}
\affiliation{Advanced Technology and Development Centre, Indian Institute of Technology Kharagpur, Kharagpur, India 721302}
\affiliation{\textit{presently at} Faculty of Materials Science \& Engineering, Technion – Israel Institute of Technology, Haifa, Israel - 3203003} 
\author{Barun Ghosh}
\affiliation{Department of Physics, Indian Institute of Technology - Kanpur, Kanpur 208016, India}
\author{Rup Kumar Chowdhury}
\affiliation{Department of Physics, Indian Institute of Technology Kharagpur,  Kharagpur, India 721302} 
\author{Amit Agarwal}
\affiliation{Department of Physics, Indian Institute of Technology - Kanpur, Kanpur 208016, India}
\author{Samit Kumar Ray}
\affiliation{Department of Physics, Indian Institute of Technology Kharagpur,  Kharagpur, India 721302}
\affiliation{S. N. Bose National Centre for Basic Sciences, Kolkata, India 700106}
\author{Debashis Chanda}
\email{Debashis.Chanda@ucf.edu}
\affiliation{NanoScience Technology Center, Department of Physics and CREOL, The College of Optics
and Photonics, University of Central Florida, Orlando, FL 32826, USA}
\affiliation{CREOL, The College of Optics and Photonics, University of Central Florida, Orlando, FL 32816}
\affiliation{Department of Physics, University of Central Florida, Orlando, FL 32816, USA}

\author{Prasanta Kumar Datta}
\email{pkdatta@phy.iitkgp.ac.in}
\affiliation{Department of Physics, Indian Institute of Technology Kharagpur,  Kharagpur, India 721302}

\title{Observation of Dynamic Screening in the Excited Exciton States in Multi-layered MoS$_2$}

\begin{abstract}
Excitonic resonance and binding energies can be altered by controlling the environmental screening of the attractive Coulomb potential. Although this screening response is often assumed to be static, the time evolution of the excitonic quasiparticles manifests a frequency-dependence in its Coulomb screening efficacy. In this letter, we investigate a ground (1s) and first excited exciton state (2s) in a multi-layered transition metal dichalcogenide (MoS$_2$) upon ultrafast photo-excitation. We explore the dynamic screening effects on the latter and show its resonance frequency is the relevant frequency at which screening from the smaller-sized 1s counterparts is effective. Our finding sheds light on new avenues of external tuning on excitonic properties.
\end{abstract}


\maketitle

Excitons or Coulomb-bound electron-hole pairs in semiconductors possess a potential for faster optical communication due to the efficient light-matter coupling and higher packing density compared to conventional electronics\cite{High229,Grosso2009}. The fundamental interaction that alters the bound state of an electron-hole pair is screening or attenuation of the Coulomb potential in presence of neighboring charge carriers which includes but not limited to atoms (dielectric screening), free carriers, excitons, and plasma. Enhanced screening leads to reduced exciton oscillator strength (OS) and binding energy (BE)\cite{PhysRevB.47.2101,PhysRevLett.115.126802}. 
A plethora of experimental studies utilize this ubiquitous phenomenon to realize external control of the excitonic states through photo-excitation\cite{Chernikov2015}, carrier-injection\cite{PhysRevLett.115.126802}, and modification of the dielectric environment\cite{Steinleitner2018,Ugeda2014,Raja2017,Castellanos_Gomez_2016}. However, the theoretical formulations that account for screening quantitatively are long-sought and well-debated\cite{PhysRev.144.708,doi:10.1002/pssb.2220850219,doi:10.1002/pssb.2220480218,doi:10.1002/pssb.201800216}.
While many of the previous studies consider long-wavelength (static) response of the environmental macroscopic polarization to account for the experimental results, static approximation largely overestimates the screening efficiency of the quasiparticles leading to inconsistencies between experimental and theoretical binding energies, oscillator strengths, Mott density etc.\cite{doi:10.1002/pssb.201800216,Sie.nanoletters} The sole reason is that static polarization responses from the surrounding charge-carrying particles are too slow to screen the excitonic Coulomb field that evolve at much faster time-scales\cite{doi:10.1002/pssb.201800216}.

The excitons are continuously annihilated and recreated through exchange interactions\cite{PhysRevB.89.205303} and scattered to free-carriers\cite{PhysRevB.32.6601,PhysRevB.32.6601,Steinhoff2017,doi:10.1002/pssb.200303153}; therefore, the Coulomb potential that binds the excitons is not static in time. Some recent theoretical studies indicate certain "characteristic frequencies" at which exciton screening is predominant\cite{PhysRevB.98.045304,doi:10.1002/pssb.201800216}.
Nevertheless, one major bottleneck in the understanding of the dynamic nature of Coulomb screening is the severe lack of experimental evidence. Therefore, the characteristic frequency (or frequencies) is an open question to date. Apart from clarity in the underlying physics, an effective device engineering necessitates comprehension of the dynamic screening effects in excitons.

Motivated by this, we explore the free-carrier and exciton-induced screening effects in excitons using ultrafast transient absorption spectroscopy that allows us to probe the temporal evolution of the screening. We choose multi-layered Molybdenum di-sulfide (MoS$_2$), a widely studied transition metal dichalcogenide (TMDC) material that offers stable, room-temperature excitons, which are also prone to sizable Coulomb screening owing to the layered architecture\cite{doi:10.1021/nn403738b}.
We track the photo-induced evolution of the ground (1s) and first excited (2s) excitonic states. Despite expected photo-bleaching like 1s state, we observe an enhancement in the 2s exciton absorption oscillator strength followed by photo-excitation.
We model the excitation-induced changes in the intrinsic dielectric permittivity (DEP) and reveal a reduction in the frequency-dependent (dynamic) dielectric permittivity of the 2s state, which triggers the enhanced absorption in the particular state. Precisely, the 2s excitons are sensitive to the intrinsic DEP dictated by 1s excitons at the 2s resonant frequency. This observation provides first-ever experimental evidence towards the perception of exciton-induced dynamic screening in semiconductors.

\begin{figure*}
\includegraphics[width=0.9\linewidth]{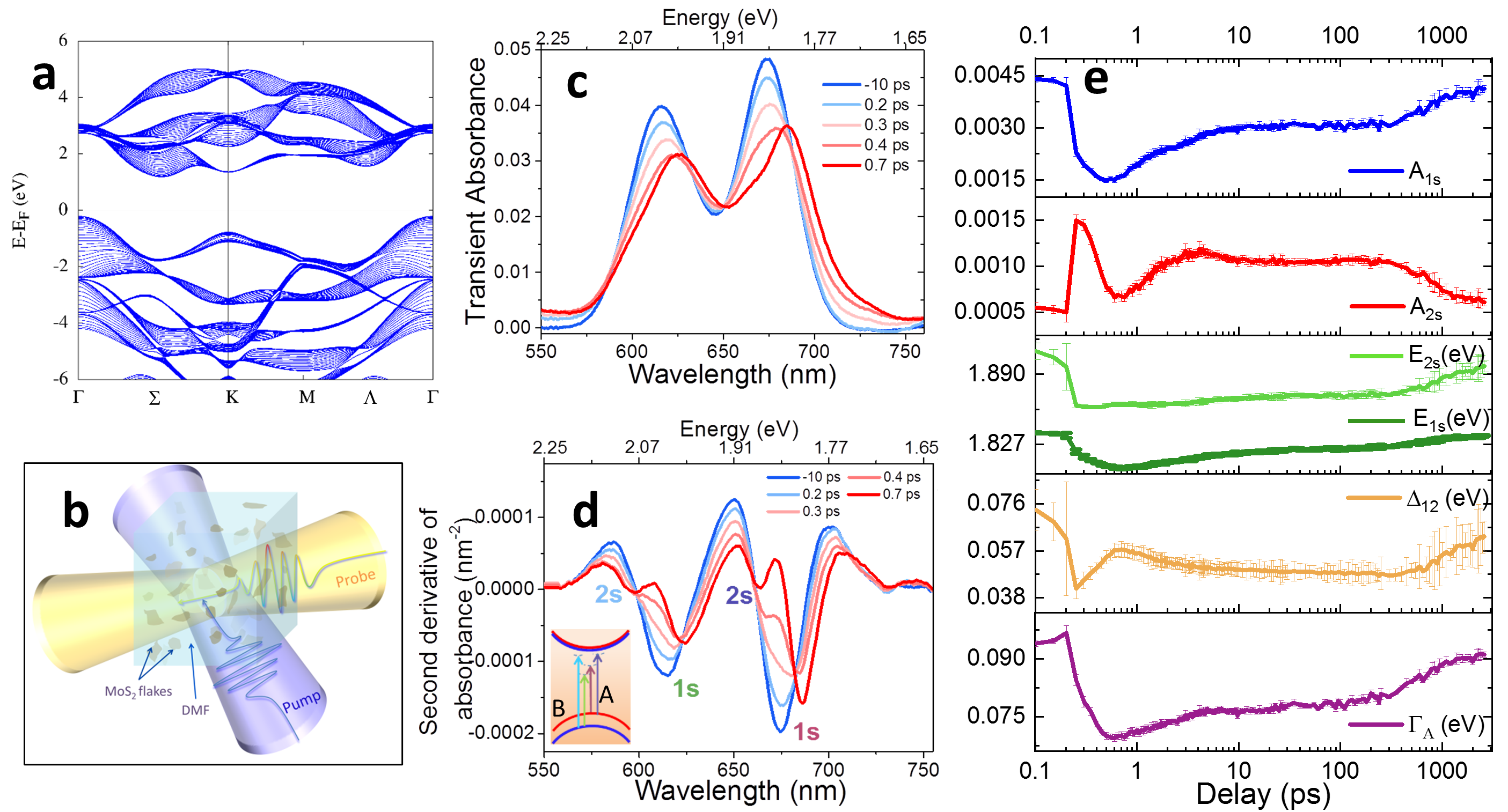}\caption{(a)Quasiparticle bandstructure of 20-layered MoS$_2$.
(b) Schematic of the multi-layered, dispersed TMDC and ultrafast light-matter interaction.
(c) Absorbance [A$_{pump}(\lambda)$] of multi-layered MoS$_2$ flakes at different probe delays for a pump-fluence of $11~\mu$J/cm$^2$, showing the A and the B exciton resonance (d) The second-derivative of $A_{\rm pump} (\lambda)$ (with respect to $\lambda$) at different probe delays.
The inset shows a schematic of the 1s and 2s excitonic states of both A and B resonances. (e) Temporal-evolution of various excitonic parameters related to ground and first excited state of A exciton, namely exciton oscillator strength, resonance energy, 1s-2s resonance energy separation, and exciton linewidth under above-bandgap pump-excitation.
\label{fig1}} 
\end{figure*}


Linear absorption spectrum of sono-chemically exfoliated multi-layered MoS$_2$ \cite{doi:10.1021/acsphotonics.5b00111} (Supplemental Material(SM) S1-S2) depicts the well-known A and B excitonic features centered around $675$ nm ($1.84$ eV) and $614$ nm ($2.02$ eV).
Quasiparticle bandstructure calculations (SM,S3) of the multi-layered TMDC using GW method is presented in Fig.~\ref{fig1}(a).

We study ultrafast transient absorption spectra of MoS$_2$  flakes obtained using a $415$ nm ($2.98$ eV) pump excitation and a CaF$_2$ generated broadband supercontinuum probe pulse and a variable pump-probe delay up to 3 ns.
A schematic illustration of the light-matter interaction is depicted in Fig.~\ref{fig1}(b).
Fig.~\ref{fig1}(c) reveals the background-corrected transient probe absorbance (SM,S5) or $A_{\rm pump}(\lambda)$, at a few selected probe delays (0.2 - 0.7 ps) and that without pump-excitation (-10 ps). With increasing probe delay, we observe red-shifted A and B absorption along with an aberrant distortion in the spectral shape of each exciton at the higher energy side.
To resolve any small spectral features, we plot the second-derivative of the absorption data in Fig.~\ref{fig1}(d). The derivative spectrum at -10 ps delay looks symmetric, whereas, at higher delays, the spectrum deviates significantly at the lower wavelength side of each exciton. This observation indicates appearance or enhancement of additional features other than the A and B ground states followed by photo-excitation.
The possibility of observing higher-order quasiparticles (trions or bi-excitons) are precluded, as they lie on the higher wavelength side of the excitonic features\cite{doi:10.1021/nl501988y}.
A possible artifact of photo-induced lattice heating is the appearance of phonon sidebands\cite{PhysRevLett.119.187402}. We perform temperature-dependent linear absorption to identify phonon sidebands at an elevated temperature (Fig. S2.2, SM). However, no asymmetry or kink appears at the higher-energy side of each exciton.
The possibility of interlayer excitons is also ruled out(SM, S11)\cite{PhysRevB.97.241404,PhysRevB.99.035443,Li2020}.
Consequently, we assign these new features to the first excited states of excitons (2s excitonic states)(also see SM,S12). We fit the derivative spectrum of the absorption data in the absence of pump and identify the 2s states of A excitons centered around $648$ nm ($1.91$ eV) having a spectral weight $\sim 1/8$ that of 1s state, as predicted \cite{PhysRev.108.1384}. From the spectral positions, we estimate exciton BE of $0.1\pm 0.007$ eV of A,1s excitons
\cite{Klingshirn2007,PhysRevLett.113.076802}.
Recent reports on bulk and multi-layered TMDC\cite{PhysRevB.97.045211,Xie:19} estimate similar values.

Temporal behavior of the various parameters including exciton oscillator strength (OS), energy resonance, linewidth, 1s-2s energy separation are found by fitting the second-order derivative (with respect to wavelength $\lambda$) of transient probe absorbance with the second-order derivative of the Gaussian-convoluted Elliott formula \cite{PhysRev.108.1384} in equation~\ref{eq1}.
\begin{equation} \label{eq1}
\frac{d^2 A_{pump}(\lambda)}{d \lambda^2} = \frac{d^2}{d \lambda^2} \left[\sum_{i = A,B} \sum_{j = 1s, 2s} \frac{A_{ij}}{\Gamma_{ij}} e^{- \frac{h^2c^2}{\Gamma_{ij}^2} 
\left(\frac{1}{\lambda}-\frac{1}{\lambda_{ij}} \right)^2}\right].
\end{equation}
Here, $A_{ij}$ is the normalized amplitude of the Gaussian (oscillator strength), $\Gamma_{ij}$ is the linewidth and $\lambda_{ij}$ is the exciton peak wavelength.
Fig.S6 in SM explicitly shows the excellent fitting of Eq.~\ref{eq1} with the data by retaining only the 1s and 2s excitons for the A and only 1s excitons for B resonances.
Extracted parameters for both ground and first excited states of A exciton for varying probe-delays are presented in Fig.~\ref{fig1}(e). 
The oscillator strength $A_{1s}$ corresponding to 1s state shows reduction suggesting Pauli-blocking and screening due to pump-induced quasiparticles\cite{doi:10.1021/acsnano.5b06488}.
We observe two exponential decay components $\tau_1$ ($\sim$ps) and $\tau_2$ ($\sim$ns) dictate the dynamics and are assigned to non-radiative carrier scattering\cite{Ceballos2016} and radiative exciton recombination\cite{doi:10.1021/nn303973r,doi:10.1021/nl503799t,Palummo2015}, respectively (SM,S7).
In contrast to the 1s state, we find an enhanced absorption OS for the 2s state, which eventually reverses to the steady-state value after a few ns.
Such absorption enhancement indicates an effective increase in BE rather than Pauli-blocking of the particular state. We will discuss this in detail.

Pump-induced charge carriers renormalize the repulsive potential energy (self-energy) and reduce the single-particle bandgap, leading to a lowering of the exciton resonance energies ($\delta_r$)\cite{doi:10.1021/acsnano.5b06488,PhysRevLett.115.126802,RevModPhys.90.021001}.
Simultaneously, the screening of the attractive interaction between the exciton constituents, reduces its BE, making it blue-shifted towards the conduction band edge: $E_j \to E_j - \delta_{r} + \delta_{b}|_{j}$ , where $E_j=\frac{hc}{\lambda_{j}}$, $j= 1s, 2s$\cite{RevModPhys.90.021001}.
For a locally-screened Coulomb potential, $\delta_b|_{1s} > \delta_b|_{2s}$\cite{Klingshirn2007}(SM,S10). Consequently, a higher red-shift in the 2s state is observed.
\begin{figure*}
\begin{center}
\includegraphics[width=1\linewidth]{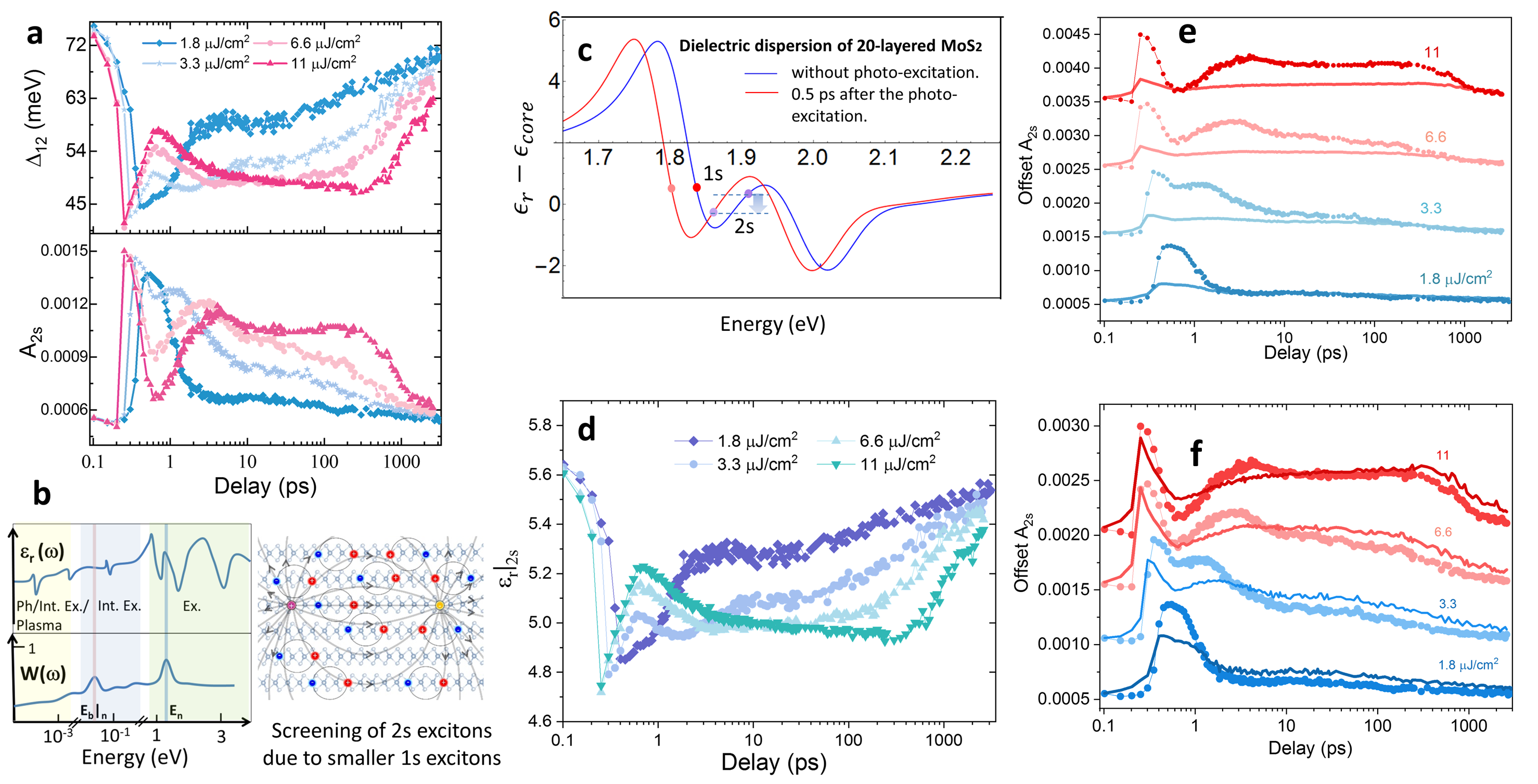}
\caption{(a) Temporal evolution of 1s-2s resonance energy separation ($\Delta_{12}$) and the absorption oscillator strength of 2s excitons (A$_{2s}$). 
(b) (Left) A pedagogical, schematic illustration of the effectiveness of the screening (W($\omega$)) and DEP ($\epsilon_r(\omega)$) of MoS$_2$ over a broad energy range.
The dielectric function below $10^{-3}$ eV is dominated by phonons (Ph), inter-excitonic transitions (Int. Ex.), and plasma, whereas excitons (Ex.) dominate the spectrum at few eV range. (Right) A schematic of the 1s (red-blue circle) and 2s excitons (purple-yellow circles). Field lines joining the latter penetrates the smaller 1s excitons which contribute to screening. 
(c)Exciton-induced permittivity of 20-layered MoS$_2$ estimated using Kramers-Kronig relations.
(d) Temporal variation of the DEP of MoS$_2$ at the 2s resonance. 
(e) Temporal variation of 2s state OS for different pump-fluences. Experimentally obtained values (same as (a)) are presented as scatter plots, and the solid lines are the estimated values from equation \ref{eq2} and \ref{eq4}. (f) Similar to (e), with minimized mean-squared error with respect to scaling of A and B, 1s absorption OS.
\label{fig2}}
\end{center}
\end{figure*}
Moreover, the excitonic linewidth is narrowed by 24 meV, followed by the pump-excitation. A plausible reason is the increase in the exciton coherence lifetime owing to Pauli-blocking of the momentum-dark states\cite{Selig2016,doi:10.1021/acs.nanolett.8b01793,Selig2016}(SM,S8).
Having an overall idea on the time-evolution of the excitonic properties, we turn to investigate the screening and 2s exciton OS enhancement.
Locally-screened, three-dimensional Hydrogen model\cite{Klingshirn2007} describes the tuning of Coulomb interaction of an electron-hole pair by employing an effective DEP experienced by the excitons for $n^{th}$ state:
\begin{equation} \label{eq2}
{\epsilon_r|_n}^2 = \frac{\mu e^4}{(4 \pi \epsilon_0)^2 2 \hbar^2  n^2 E_b|_n }.
\end{equation}
Where, $E_b|_{n}$ is exciton BE of $n^{th}$ exciton state, $\mu$ is exciton effective mass and $\epsilon_r|_n$ is effective dielectric constant.
This quantity $\epsilon$ summarizes all Coulomb screening effects experienced by an electron-hole pair\cite{PhysRevLett.113.076802}. If we consider an oversimplified picture of an exciton, where an electron and hole pair is a static entity in real space and time, the static DEP of the environment will describe the screening interactions sufficiently. 
However, different many-body interactions take place that continuously abolish and create the Coulomb pairs. For instance, excitons get ionized to free carriers through phonon-interactions, annihilated and recreated through exchange interactions, scattered to momentum-dark states upon phonon-interactions, etc. Each of these processes have their specific time-scales.
For example, exchange interaction happens at the frequency of the excitonic resonance; exciton ionization happens at the frequency related to the BE of the excitons. Accordingly, the medium surrounding an exciton screens the electron-hole interactions at those particular frequencies. Polarizability response of the environment at other frequencies are either too slow or too fast to affect the Coulomb interactions significantly.

In the case of an ideal two-dimensional semiconductor, the environmental screening of an exciton includes the substrate-response and a minimal response from the semiconducting layer itself. For a three-dimensional system like the one we study, the field-lines joining an electron and a hole are essentially within the same material. Hence, the screening of the Coulomb field is predominantly from the surrounding layers.

In our experiments, pump-induced excitons and carriers modify the charge environment of the 2s excitons such that its effective permittivity is reduced and OS is enhanced. Although we consider the dynamic screening in the TMDC, it is almost an improbable task to determine the effectiveness of the screening at each frequency of the electromagnetic spectrum.
However, as we discussed earlier, some particular frequencies (resonance frequency, binding energy/$\hbar$) are more effective than the other. In Fig.~\ref{fig2}(b), we sketch the frequency-dependent effectiveness (W($\omega$)) and the environmental (intrinsic) DEP for pedagogical purpose; we estimate the effective permittivity of an excitonic state by
\begin{equation} \label{eq3}
\epsilon_r|_n = \frac{\int_{0}^{\infty} W(\omega)\epsilon_r(\omega) d\omega}{\int_{0}^{\infty} W(\omega) d\omega}
\end{equation}
It is evident that static dielectric permittivity increases due to the pump-induced elevated population of charge carriers.
Also, the plasma frequencies owing to the carrier injection is estimated at $\sim$ $10^{-2}$ meV, much less than the 2s exciton BE.
Therefore, the effective permittivity of 2s states reduces neither at static limits nor at the binding energy range. The remaining frequency of interest is the resonance of the 2s state.

We present the delay-dependent 1s-2s resonance energy separation ($\Delta_{12}$) and $A_{2s}$ in Fig.~\ref{fig2}(a) for four different pump-fluences.
Notably, the non-monotonic dynamics of $A_{2s}$ and $\Delta_{12}$  seem to be inversely correlated, with the maxima in $A_{2s}$ temporally coinciding with the minima in $\Delta_{12}$. This observation 
reaffirms the existence of strong dynamic screening. Consequently, we trace the time-dependent intrinsic DEP of MoS$_2$ at the 2s resonance energy.

Following equation \ref{eq2} and calculated exciton reduced mass of $0.24m_0$, we extract the effective steady-state DEP of $5.7$.
We separate the effective DEP of the 2s excitons into two components-(i)permittivity at exciton resonance due to 1s excitons (ii) a core DEP due to other interband transitions as a cumulative effect from all frequencies other than the resonance. While the first term evolves with photo-excitation, the other is assumed to be fixed. This assumption serves our purpose of investigating the variation of screening at the resonance.

We employ Kramers-Kronig (KK) relation\cite{doi:10.1142/7184} based on the linear absorbance data in the visible region (SM, S9) to find out (i), i.e., the real part of the dielectric dispersion ($\epsilon_{r,_{KK}}$($\lambda$)=$\epsilon_r-\epsilon_{core}$) due to excitons.
Note that material permittivity is given by $\epsilon_{r,_{KK}}(\lambda)+\epsilon_{core}$, where $\epsilon_{core}$ is a nearly frequency-invariant background permittivity originating from interband transitions. As we use TMDC flakes in dispersion, with probe beam size ($\sim 80~\mu$m beam-diameter) being few-orders larger than the individual suspended flakes, the experimentally measured absorbance is less than the actual absorbance of a 20-layered MoS$_2$ flake.
Therefore, we estimate a scaling parameter ($5\pm2$) of the absorbance by extrapolating an earlier work\cite{Castellanos_Gomez_2016} (see SM,S9).
Next, we plot the exciton-induced DEP with (for probe-delay of 0.5 ps) and without photo-excitation in Fig.~\ref{fig2}(c) using the corresponding experimentally obtained $\Delta_{12}$ values in KK equation.
As compared to the un-pumped condition, a reduction in the DEP at the 2s exciton resonance is comprehensible.
We note that the exciton OS ($A_j$) in a three dimensional semiconductor varies as $A_{j} \propto r_{j}^{-3}$, where $r_j$ denotes the exciton radius\cite{doi:10.1142/7184}.
The exciton radius, in turn, varies linearly with the effective DEP\cite{gaponenko_2010}, leading to
\begin{equation} \label{eq4}
A_{j} \propto \epsilon_r^{-3}.
\end{equation}
Thereafter, we calculate the delay-dependent reduction in $\epsilon_r|_{2s}$ and accordingly find A$_{2s}$ at different probe delay (Fig.\ref{fig2} (d) and (e)).
\begin{figure}
  \begin{center}
    \includegraphics[width=0.47\textwidth]{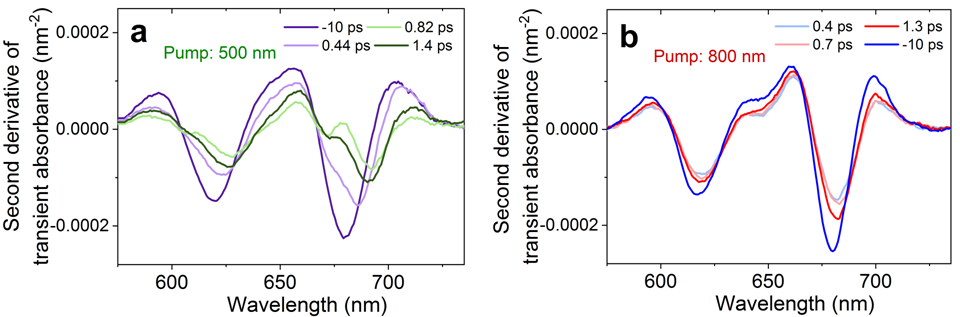}
    \caption{Second-order derivative of the TA spectra at a few selected delays after photo-excitation with (a) 500 nm and (b) 800 nm pump of fluence of 6 $\mu$J/cm$^2$ and 264 $\mu$J/cm$^2$, respectively.}
  \end{center}
  \label{fig3}
\end{figure}
Remarkably, the temporal evolution of the $A_{2s}$ is well-reproduced qualitatively. However, calculated values underestimate experimental A$_{2s} $, almost by a factor of $\frac{1}{2}$. This is improved by optimizing the scaling parameter by 40 $\%$, which lies within the standard error of the mentioned data extrapolation.
Re-calculated A$_{2s}$ values displayed in Fig.~\ref{fig2}(f) depicts a better quantitative estimation of the experimental data.
This observation excellently demonstrates the screening of 2s excitons due to the smaller-sized 1s excitons. Photo-excitation triggers reduced effective permittivity due to reduced 1s-2s energy separation, which facilitates reduced screening or "antiscreening"\cite{Brink_2000} leading to the enhanced exciton OS. This effect has an upper-limit pertaining to the 2s exciton resonance entering anomalous dispersion region of 1s exciton oscillator(SM,S10).

We repeat the pump-probe experiments with thin-films of multi-layered flakes on quartz substrates. We excite the sample linearly with 500 nm and 800 nm pump and observe the resulting second-derivative of the transient absorption in Fig.3.
The 800 nm pumping condition constitutes an interesting case where A,1s excitons do not form.
Therefore, in contrast to the 415 nm and 500 nm pumping, the kinks owing to the enhanced 2s excitonic features do not appear for 800 nm pump and thereby re-affirms the role of 1s excitons in 2s exciton screening.

In this letter, we provide a strong experimental evidence of dynamic Coulomb screening in excitons and reveal that environmental polarization response at its resonance effectively controls the screening and hence the absorption strength of the excitons.
We observe that despite the presence of photo-induced carriers that induce increased static-screening, 2s exciton absorption is enhanced.
We reproduce this aberrant experimental observation by considering the intrinsic dielectric permittivity at its resonance owing to the 1s excitons. Thus, it reveals that the exciton resonance frequency is the characteristic frequency at which other polarizations effectively screen the Coulomb interactions. This observation opens up new opportunities to tailor exciton resonances without losing the population, unlike existing approaches.

\begin{acknowledgements}
M.K. thanks Dr. Simone Peli for helpful discussions. The authors gratefully acknowledge SDGRI-UPM project of IIT Kharagpur for all the necessary equipment.The authors thank Mr. Abhinav Kala and Prof. Venu Gopal Achanta from Tata Institute of Fundamental Research, India for providing a low-temperature absorption measurement of few-layered MoS$_2$ sample.
\end{acknowledgements}

\end{document}